\documentclass{article}

\usepackage{amsmath,amsfonts,amssymb,bm}
\usepackage[margin=1in]{geometry}
\usepackage{bbm}
\usepackage{graphicx}
\usepackage{authblk}
\usepackage[round]{natbib}

\usepackage{url} %this package should fix any errors with URLs in refs.

\begin{document}

%% ------------------------------------------------------------------------ %%
%  Title
%
% (A title should be specific, informative, and brief. Use
% abbreviations only if they are defined in the abstract. Titles that
% start with general keywords then specific terms are optimized in
% searches)
%
%% ------------------------------------------------------------------------ %%

\title{Revealing the statistics of extreme events hidden in short weather forecast data}

\author[1]{Justin Finkel \thanks{ju26596@mit.edu}}
\author[2]{Edwin P. Gerber}
\author[3]{Dorian S. Abbot}
\author[2]{Jonathan Weare}

\affil[1]{Department of Earth, Atmospheric, and Planetary Sciences, Massachusetts Institute of Technology}
\affil[2]{Courant Institute of Mathematical Sciences, New York University}
\affil[3]{Department of Geophysical Sciences, University of Chicago}

\maketitle

\section*{Key points}
\begin{enumerate}
    \item Extreme weather risk is inherently difficult to quantify because of data scarcity. 
    \item Subseasonal weather forecast ensembles are an untapped resource for computing statistics of extremes, done here by statistical weighting.
    \item We characterize stratospheric extremes up to once in 500-year events from 46-day hindcast ensembles across 20 winters.
\end{enumerate}

%% ------------------------------------------------------------------------ %%
%
%  ABSTRACT and PLAIN LANGUAGE SUMMARY
%
% A good Abstract will begin with a short description of the problem
% being addressed, briefly describe the new data or analyses, then
% briefly states the main conclusion(s) and how they are supported and
% uncertainties.

% The Plain Language Summary should be written for a broad audience,
% including journalists and the science-interested public, that will not have 
% a background in your field.
%
% A Plain Language Summary is required in GRL, JGR: Planets, JGR: Biogeosciences,
% JGR: Oceans, G-Cubed, Reviews of Geophysics, and JAMES.
% see http://sharingscience.agu.org/creating-plain-language-summary/)
%
%% ------------------------------------------------------------------------ %%

%% \begin{abstract} starts the second page

\begin{abstract}

Extreme weather events have significant consequences, dominating the impact of climate on society. While high-resolution weather models can forecast many types of extreme events on synoptic timescales, long-term climatological risk assessment is an altogether different problem. A once-in-a-century event takes, on average, 100 years of simulation time to appear just once, far beyond the typical integration length of a weather forecast model. Therefore, this task is left to cheaper, but less accurate, low-resolution or statistical models. But there is untapped potential in weather model output: despite being short in duration, weather forecast ensembles are produced multiple times a week. Integrations are launched with independent perturbations, causing them to spread apart over time and broadly sample phase space. Collectively, these integrations add up to thousands of years of data. We establish methods to extract climatological information from these short weather simulations. Using ensemble hindcasts by the European Center for Medium-range Weather Forecasting (ECMWF) archived in the subseasonal-to-seasonal (S2S) database, we characterize sudden stratospheric warming (SSW) events with  multi-centennial return times. Consistent results are found between alternative methods, including basic counting strategies and Markov state modeling. By carefully combining trajectories together, we obtain estimates of SSW frequencies and their seasonal distributions that are consistent with reanalysis-derived estimates for moderately rare events, but with much tighter uncertainty bounds, and which can be extended to events of unprecedented severity that have not yet been observed historically. These methods hold potential for assessing extreme events throughout the climate system, beyond this example of stratospheric extremes.

\end{abstract}

\section*{Plain Language Summary}
Weather extremes are a continually recurring threat to human life, infrastructure, and economies. Yet, we only have sparse datasets of extremes, both simulated and observed, because by definition they occur rarely. We introduce an approach to extract reliable extreme event statistics from a non-traditional data source: short, high-resolution weather simulations. With only 20 years of 46-day weather forecasts, we estimate the magnitudes of once-in-500-year events. 
%% ------------------------------------------------------------------------ %%
%
%  TEXT
%
%% ------------------------------------------------------------------------ %%

%%% Suggested section heads:
% \section{Introduction}
%
% The main text should start with an introduction. Except for short
% manuscripts (such as comments and replies), the text should be divided
% into sections, each with its own heading.

% Headings should be sentence fragments and do not begin with a
% lowercase letter or number. Examples of good headings are:

% \section{Materials and Methods}
% Here is text on Materials and Methods.
%
% \subsection{A descriptive heading about methods}
% More about Methods.
%
% \section{Data} (Or section title might be a descriptive heading about data)
%
% \section{Results} (Or section title might be a descriptive heading about the
% results)
%
% \section{Conclusions}

\section{Introduction}
\label{sec:intro}

The atmosphere's extreme, irregular behavior is, in some ways, more important to characterize than its typical climatology. A society optimized for average historical weather patterns is highly exposed to damage from extreme heat and cold, flooding, and other natural hazards. Moreover, extremes may respond more sensitively than mean behavior to climate change, an argument supported by elementary statistics \cite{Wigley2009effect}, empirical observations \cite{Coumou2012decade,AghaKouchak2014global,OGorman2012sensitivity,Huntingford2014potential,Naveau2020statistical} and simulations \cite{Pfahl2017understanding,Myhre2019frequency}. Recent unprecedented extreme weather events demonstrate the serious human impacts \cite{Mishra2018hydroclimatological,Van2017attribution,Goss2020climate,Fischer2021increasing}. The overall ``climate sensitivity'' \cite{Hansen1984climate}, summarized by a change in global-mean temperature, does not do justice to these consequences, which has led to the development of ``event-based storylines'' \cite{Shepherd2018storylines,Sillmann2021event} as a more tangible expression of climate risk. 

The intermittency of extreme events makes precise risk assessment exceedingly difficult. 100 flips of a biased coin with $\mathbb{P}\{\text{Heads}\}=0.01$ is almost as likely to yield zero heads (probability 0.366) as one head (probability 0.370), and half as likely to yield two heads (probability 0.185). Similarly, in a 100-year climate simulation or historical record, a once-per-century event will more likely appear either non-existent or twice as likely as it really is. This difficulty is present for a stationary climate, but worsens in the presence of time-dependent forcing, anthropogenic or otherwise. The limited historical record forces us to use numerical models as approximations, introducing a dilemma: we can run cheap, coarse-resolution models for long integrations, providing reliable statistics of a biased system, or expensive, high-resolution models for short integrations, which have lower bias but provide statistics with higher variance due to under-sampling. For example, the Integrated Forecast System (IFS) of the European Center for Medium-Range Weather Forecasts (ECMWF) is one of the most accurate weather models available today, running at high resolutions of $\sim$16-32 km \cite{ECMWF2016ifs5ensemble}. The forecasts are skillful, but typically last for a single season or less---far too short a duration to estimate rare event probabilities directly. 

However, these forecasts are launched multiple times every week in large parallel ensembles, which can be exploited to bridge the gap from weather to climate timescales. The key is to include the data from ensemble members in a statistically principled way. Our main contribution in this paper is to introduce methods to achieve this, using the ensemble forecasts archived in the subseasonal-to-seasonal (S2S) project at ECMWF \cite{Vitart2017subseasonal}.

Specifically, in this work we estimate probabilities of sudden stratospheric warming (SSW) events, in which the winter stratospheric polar vortex rapidly breaks down from its typical state with a strong cyclonic circulation over the winter-hemisphere pole. The associated subsidence of air in the polar stratosphere leads to adiabatic warming, causing lower-stratospheric temperatures to rise up to 40 K or more over a few days \cite{Baldwin2021sudden}. The breakdown of the stratospheric vortex %winds forces upward-propagating planetary waves to break at lower and lower levels, ultimately impacting synoptic scale waves to 
exerts a ``downward influence'' on tropospheric circulation \cite{Baldwin2001stratospheric,Baldwin2003stratospheric,Hitchcock2014downward,Kidston2015stratospheric}. The midlatitude jet and storm track shift equatorward, bringing extreme cold spells and other anomalous weather to nearby regions \cite{Kolstad2010association,Kretschmer2018different}. For example, \cite{King2019observed} documents the impact of an SSW on extreme winter weather over the British Isles, the so-called ``Beast from the East'' in February 2018. SSWs are a demonstrated source of surface weather predictability on the subseasonal-to-seasonal (S2S) timescale \cite{Sigmond2013enhanced,Butler2019subseasonal,Scaife2022longrange}. Pushing this ``frontier'' of weather forecasting can improve disaster preparation and resource management in the face of meteorological extremes \cite{White2017potential,Bloomfield2021subseasonal}. For these reasons, there is keen interest in improving (i) the prediction of SSW itself beyond the horizon of $\sim$10 days that marks the current state-of-the-art \cite{Tripathi2016examining,Domeisen2020role}, and (ii) understanding of the long-term frequency and other climatological statistics of SSWs \cite{Butler2015defining, Gerber_etal2022}.  
% TODO paper summary 

\section{Data and definitions}
\label{sec:datadef}

\begin{figure}%[tbhp]
	\centering
	\includegraphics[width=0.98\linewidth,trim={0cm 0cm 0cm 0cm},clip]{"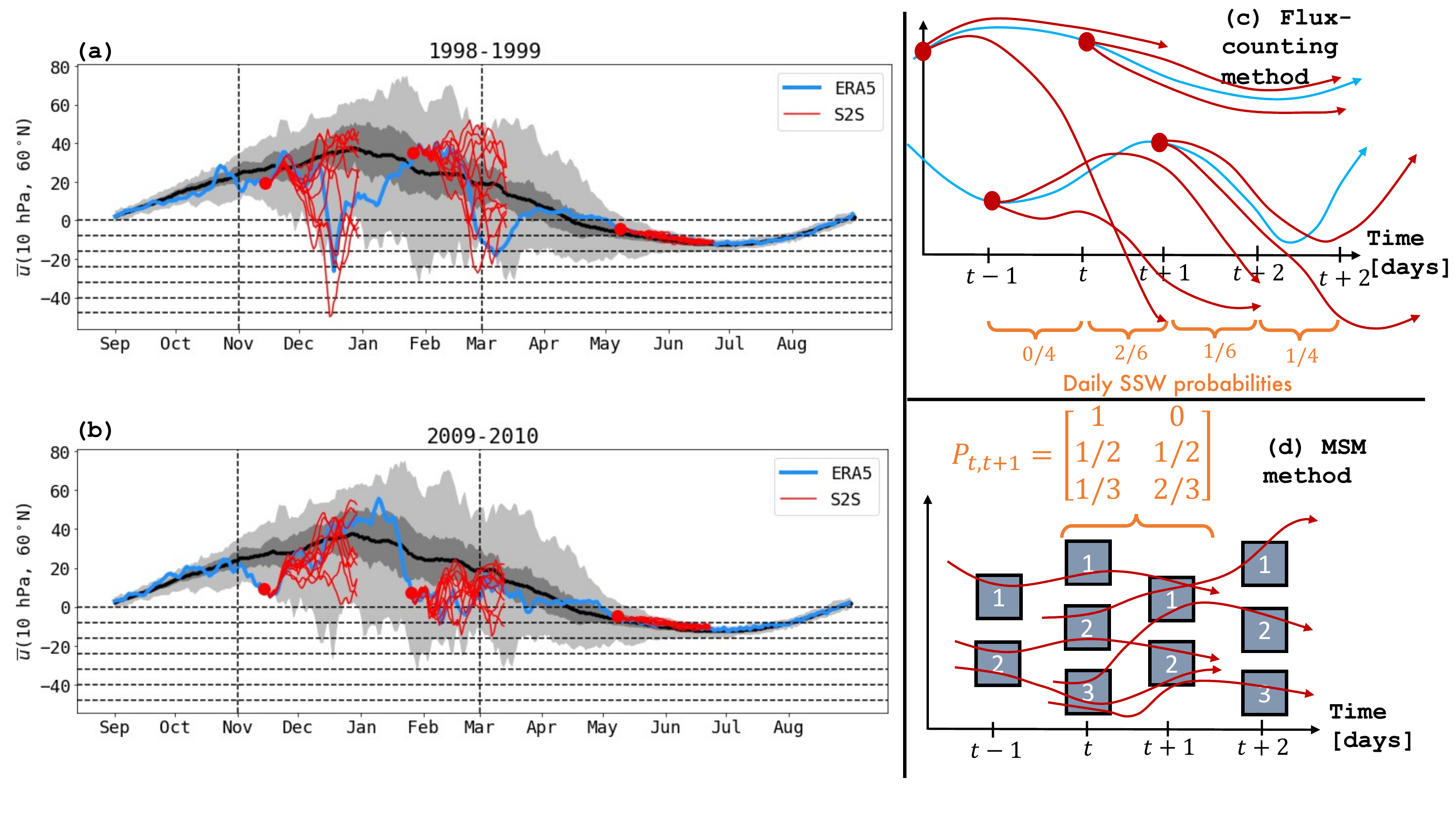"}
    \caption{\textbf{Climatology of polar vortex and illustration of dataset}. Black curves show the mean seasonal cycle of $\overline{u}$(10 hPa, 60$^\circ$N), abbreviated as $U$, and two gray envelopes show the percentile ranges 25-75 and 0-100, respectively. All statistics are computed with respect to the 61-year ERA-5 dataset between 1959 and 2019. Two individual years are shown in blue: 1998-1999 (a) and 2009-2010 (b). Two ensembles of S2S hindcasts (red) are shown each winter, a small sample from the large S2S dataset of two ensembles \emph{per week} from the ECMWF IFS. Horizontal dashed lines mark several different SSW thresholds $U^{\mathrm{(th)}}$ used in this study, including the standard threshold of 0 m/s and several more extreme ones. The time window Nov. 1 - Feb. 28 is marked by vertical dashed lines. When $U$ crosses $U^{\mathrm{(th)}}$ downward for the first time within the time window, an SSW has occurred. (c): schematic of the flux-counting method. (d): schematic of the Markov State Model (MSM) method. }
	\label{fig:ssw_data}
\end{figure}

Fig. \ref{fig:ssw_data}(a,b) show the evolution of zonal-mean zonal wind at 10 hPa and 60$^\circ$N, a standard index for the strength of the stratospheric polar vortex \cite{Butler2019subseasonal}, which we abbreviate $U$. Blue timeseries show $U$ through two consecutive winters: (a) 1998-1999, in which an extreme SSW occurred as quantified by the deep drop in $U$ in mid-December, and (b) 2009-2010, when a more mild SSW occurred in February. Both timeseries are superimposed upon the 1959-2019 ERA-5 climatology. %A black curve shows the daily mean climatology, the dark gray envelope shows the interquartile range, and the lighter envelope represents the min-max range on each day of the annual cycle.

$U$ is typically positive throughout the winter months, characterizing a strong circumpolar jet that forms in the stratosphere during the polar night. %Occasionally, however, the vortex breaks down and $U$ reverses direction, becoming negative in the middle of winter. 
The standard definition of an SSW event is that $U$ changes sign \cite{Butler2015defining}, but it does not capture the range of intensities between events. Clearly, December 1998 exhibited a much stronger breakdown of the vortex than February 2010. More intense SSW events have been linked to stronger tropospheric impacts \cite{Karpechko2017predictability,Baldwin2021sudden}, which motivates our efforts to distinguish between them. Historical data can provide reasonably robust estimates of moderately rare events such as February 2010, in which $U$ barely reversed sign; events of this magnitude occur on average every two years. On the other hand, extraordinary events like December 1998 have only been observed a few times. %are poorly constrained due to small sample size. 

We define an SSW as the first decrease in $U$ below a threshold $U^{(\mathrm{th})}$ during the ``SSW season'' of Nov. 1-Feb. 28. We only count the first event of a winter to exclude the subsequent oscillations of $U$ about $U^{\text{(th)}}$ as separate SSW events, without complicating the definition with a minimum separation time as in \cite{Charlton2007part1}. The main quantity of interest is the \emph{rate}: the average number of SSW events per year, a number between zero and one. Equivalently, the reciprocal of the rate is called the \emph{return period}: the expected number of years to wait before an event of a given severity. For the standard threshold $U^{(\mathrm{th})}=0$, the rate is approximately $0.6$ \cite{Baldwin2021sudden}, but we will consider a range of severities by varying $U^{(\mathrm{th})}$ down to $-52$ m/s. 

\begin{figure}%[tbhp]
	\centering
	\includegraphics[trim={0cm 0cm 0cm 0cm},clip,width=.99\linewidth]{"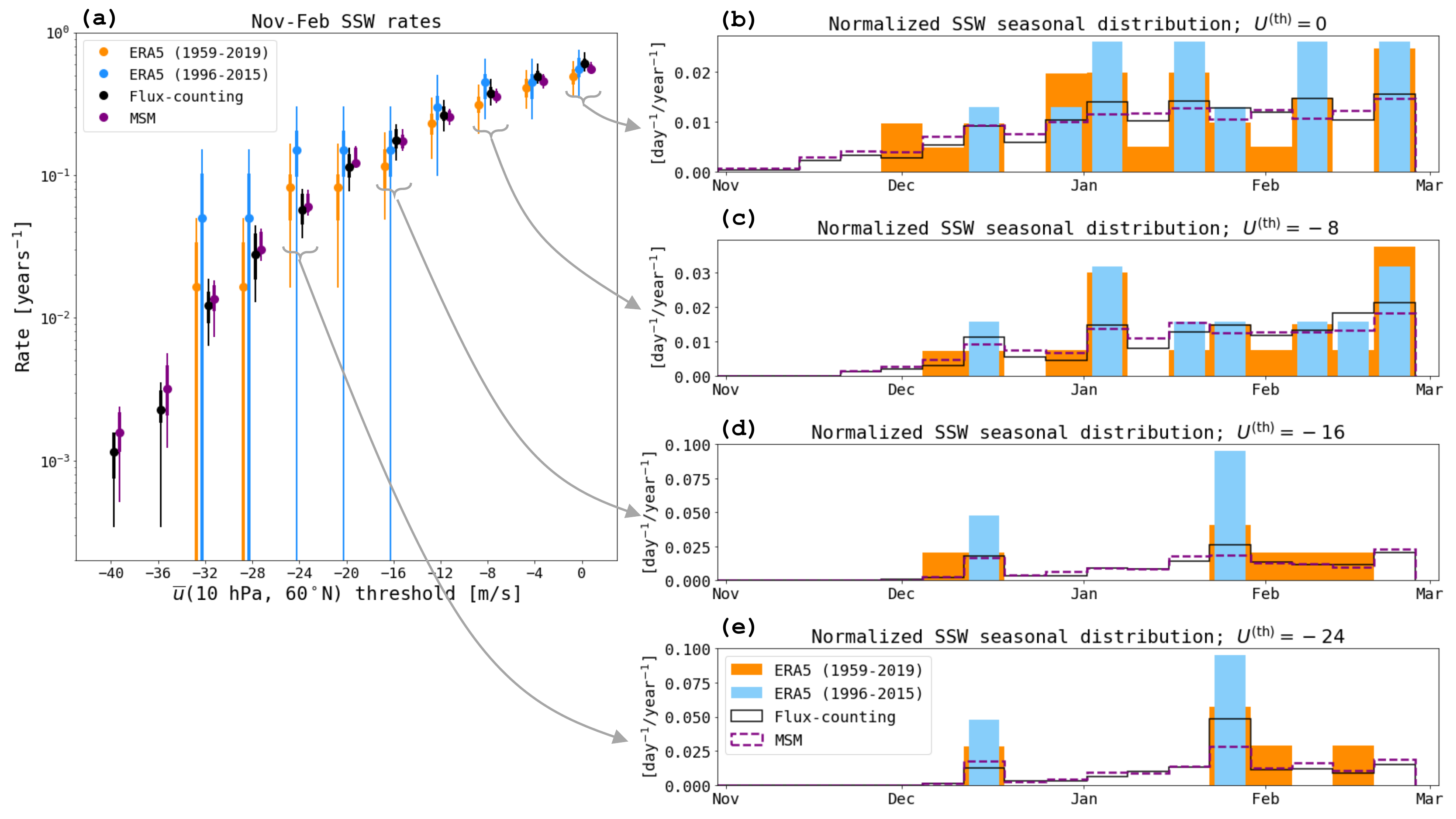"}
	\caption{\textbf{Rate estimates derived from S2S and reanalysis}. Left: SSW rate (inverse annual probability of SSW) as a function of zonal wind threshold, $U^{(\mathrm{th})}$, estimated by the four methods described in the text. Error bars indicate 95\% confidence intervals. MSM and flux-counting error bars are computed by bootstrapping on entire years of data. ERA5 error bars are computed  analytically as the 2.5-97.5 percentile range of the success rate of a binomial random variable with a success probability given by the estimated rate and a number of trials given by the number of years in the record (20 or 61). Error bars going off the bottom of the plot include zero (note the log scale). Right: seasonal distribution of SSW events at four selected thresholds, according to each of four methods. All histograms have a bin width of 7 days and are rescaled to have unit area.}
	\label{fig:rates}
\end{figure}

One can estimate the rate with reanalysis by counting the fraction of years with an SSW event. Fig.~\ref{fig:rates}a shows two reanalysis-derived rate estimates as a function of $U^{\text{(th)}}$: the blue points use 61 years of data (1959-2019) while the orange points use only 20 years of data (1996-2015). The corresponding error bars encompass the 50\% (thick lines) and 95\% (thin lines) confidence intervals of $(X_1+\hdots+X_n)/n$, where the $X_i$'s are independent Bernoulli random variables with success rate given by the estimated rate, and a number of trials equal to $20$ (blue) or $61$ (orange). Fig.~\ref{fig:rates}b-e shows the corresponding seasonal distribution of events at four selected thresholds, with histograms normalized to have unit area. It may appear inconsistent that the support of the distribution at $U^{\mathrm{(th)}}=-8$ is not fully contained in the support of the distribution at $U^{\mathrm{(th)}}=0$; for example, the third blue and orange bins of February have positive weight in panel (c), but zero weight in panel (b). There is no contradiction: although every winter with an SSW at level $-8$ also must have an SSW at level $0$, the weaker threshold is crossed first and is sometimes counted in previous weeks.

%Overall, the shorter period (1996-2015) suggests slightly higher rates across the board, due to an outsized share of the most extreme events such as Dec. 1998. The recent heightened rates have been documented before, and attributed to multi-decadal cycles such as the quasi-biennial oscillation (QBO), El Ni{\~n}o southern oscillation (ENSO), and Atlantic meridional overturning circulation (AMOC) \cite{Reichler2012stratospheric,DimdoreMiles2021origins}. The spate of recent SSWs may also explain some increasing cold-weather outbreaks despite an overall warming planet \cite{Kretschmer2018more,Garfinkel2017stratospheric}. However, the error bars in Fig.~\ref{fig:rates} remind us that mere random sampling may explain a lot of the difference in SSW rates between periods. Shrinking the error bars would help to confirm or refute any hypotheses about changing SSW rates. 

The other estimates displayed in Fig.~\ref{fig:rates} are derived from the S2S dataset, which consists of hindcast trajectories launched in ten-member perturbed ensembles (plus a control member that we omit from our analysis). We use only hindcast data produced by the 2017 version of the ECMWF IFS: that is, integrations initialized from past initial conditions for the 20 years prior, in our case from autumn 1996 to spring 2016 (labeled 1996-2015 in the plots). Each ensemble member has small perturbations applied to its initial conditions, and is integrated forward with stochastically perturbed tendencies \cite{Buizza1999stochastic,Berner2009spectral}. For details on the the model, see \cite{Vitart2017subseasonal} and \cite{ECMWF2016ifs5ensemble}. The dataset is publically accessible at \url{https://apps.ecmwf.int/datasets/data/s2s/}.

The total number of days contained therein is roughly
%\begin{linenomath*}
\begin{align}
20\text{ years}\times\frac{52\text{ weeks}}{\text{ year}}\times\frac{2\text{ ensembles}}{\text{week}}\times\frac{10\text{ members}}{\text{ensemble}}\times\frac{47\text{ days}}{\text{member}}\approx2700\text{ years}
\end{align}
%\end{linenomath*}
or, only counting Nov-Feb, 900 years. Many of these extra ensemble members reach farther into the negative-$U$ tails than the reanalysis. Thinking of these as alternative realities, we can calculate otherwise inaccessible probabilities. %, as outlined in the next section.

\section{Two estimates of long return times from short trajectories}
\label{sec:longshort}

To take advantage of the S2S data, we have to overcome two complications. First, not all trajectories are independently sampled: on the contrary, all members of an ensemble are initialized close to reanalysis, and take several days to separate. Thus, the effective sample size is smaller than 900 years. Second, no individual ensemble can directly provide an SSW probability beyond the 46-day time horizon, which is well short of the 120 days between November 1 and February 28 when SSWs are allowed to happen. We cannot use hindcasts directly to estimate the rate, because we need to know what would have unfolded if the 46-day simulation were to continue. The challenge is to make use of the ``hanging'' trajectory endpoints, such as the eight members of the first ensemble shown in Fig.~\ref{fig:ssw_data} which do not dip below the threshold. Below, we present two related, but distinct methods: flux-counting and Markov state modeling.  

\subsection{Flux-counting for direct estimates}
\label{sec:fluxcounting}
The first approach is quite simple, as sketched in Fig. \ref{fig:rates}c: we compute the probability of an SSW on each day by calculating the fraction of trajectories that cross the threshold on that day (avoiding double counting by keeping track of ``active" trajectories as detailed below), and the sum up all the daily probabilities over the season. Formally, we decompose the winter months of interest into a sequence of one-day windows, which is the sampling resolution of S2S:
%\begin{linenomath*}
\begin{align}
	T_0=\text{Nov 1}, T_0+1=\text{Nov. 2},\hdots,T_1=\text{Feb. 28}
\end{align}
%\end{linenomath*}
and estimate the probability of an SSW separately for each calendar day. By our definition, an SSW can happen at most once per season, to ensure the events are disjoint and have additive probabilities:
%\begin{linenomath*}
\begin{align}
	\text{Rate}&=\sum_{t=T_0}^{T_1}\mathbb{P}\{\text{SSW on day }t\}\\
	&=\sum_{t=T_0}^{T_1}\mathbb{P}\Big\{\min_{T_0\leq s<t}U(s)>U^{\text{(th)}}\text{ and }U(t)\leq U^{\text{(th)}}\Big\}
\end{align}
%\end{linenomath*}
% TODO: possibly change T_1 to T_f

The summand can be considered a probability per day of crossing the threshold $U^{(\mathrm{th})}$, i.e., one of the horizontal dashed lines in Fig.~\ref{fig:ssw_data}a,c. It is estimated by averaging over all hindcast trajectories that are ``active'' on calendar day $t$, meaning those launched some day between $t-46$ and $t$. More precisely, if we enumerate the active trajectories by $i\in\mathcal{I}(t)=\{1,...,N(t)\}$ and denote the $i$'th trajectory's zonal wind by $U_i$, then the estimate of daily SSW probability is
%\begin{linenomath*}
\begin{align}
	\label{eq:ed}
	\mathbb{P}\{\text{SSW on day }t\}=\frac{1}{N(t)}\sum_{i\in\mathcal{I}(t)}\mathbb{I}\Big\{\min_{T_0\leq s<t}U_{i}(s)>U^{\text{(th)}}\Big\}\mathbb{I}\Big\{U_{i}(t)\leq U^{\text{(th)}}\Big\}
\end{align}
%\end{linenomath*}
where $\mathbb{I}$ is an indicator function, equal to 1 if the argument is true and 0 if the argument is false. In words, we count the trajectories that dip below $U^{\text{(th)}}$ for the first time on day $t$, as a fraction of all trajectories that are active on that day. The past of ensemble member $i$ before its initialization date is given by the corresponding reanalysis from which it branched. 

Summing up these probabilities from Nov. 1 to Feb. 28, and sweeping over all thresholds $U^{\text{(th)}}$, we obtain the black curve in Fig.~\ref{fig:rates}a. Error bars come from a bootstrapping procedure: we apply the estimate~\eqref{eq:ed} to 20 different random 10-year subsets of $\{1996,\hdots,2015\}$, calculate the 2.5th and 97.5th percentiles of rate estimates, and form the pivotal 95\% bootstrap confidence interval (see \cite{Wasserman2004all}, chapter 8, for a formal account, although we have modified the procedure by sampling without replacement to maintain independence of different years.)

The average~\eqref{eq:ed} is a sum of \emph{dependent} random variables, with all ensemble members in a given year sharing common history. This increases the variance of the estimator or, in other words, reduces the effective sample size from 900 years. This situation is common in the Monte Carlo simulation for inverse problems. But the error bars make clear that flux-counting enjoys a tremendous advantage over the direct ERA5 estimate. At all thresholds, the flux-counting error bar overlaps with  the ERA5 error bar, but is much smaller. This gives us confidence to trust the flux-counting estimate farther into the tail where no ERA5 data are available. 

\subsection{Markov state model}
\label{sec:msm}

The second method is more intricate, but delivers more insight into the predictability of SSWs. We construct a \emph{Markov state model} (MSM) \cite{Deuflhard1999MSM,Pande2010MSM,Chodera2014msm} which is sketched in Fig. \ref{fig:rates}d. On each day $t$, we partition state space into a disjoint collection of bins $S_{t,1},S_{t,2},\hdots,S_{t,M_t}$ and approximate the transition probability matrix for each time-step from $t$ to $t+1$,
%\begin{linenomath*}
\begin{align}
	\label{eq:transition_matrix}
	P_{t,t+1}(j,k)&=\mathbb{P}\{\mathbf{X}(t+1)\in S_{t+1,k}|\mathbf{X}(t)\in S_{t,j}\},
\end{align}
%\end{linenomath*}
by counting the transitions between corresponding boxes. Explicitly,
%\begin{linenomath*}
\begin{align}
	P_{t,t+1}(j,k)&=\frac{\sum_{i\in\mathcal{I}(t)}\mathbb{I}\{\mathbf{X}_{i}(t)\in S_{t,j}\}\mathbb{I}\{\mathbf{X}_{i}(t+1)\in S_{t+1,k}\}}{\sum_{i\in\mathcal{I}(t)}\mathbb{I}\{\mathbf{X}_{i}(t)\in S_{t,j}\}}
\end{align}
%\end{linenomath*}

 The matrices are row-normalized, which corrects for the redundancy and statistical dependence between ensemble members. This sequence of matrices is the key ingredient that enables all downstream calculations. Choosing the partition of state space is a crucial step which involves a tradeoff: too few clusters will coarsen the dynamics too much, whereas too many clusters will reduce the number of data points in each cluster and thus increase the statistical noise involved in estimating $P_{t,t+1}(j,k)$. There is a lack of general theory on how to construct MSMs, but here we exploit the particular structure of the dataset to validate our algorithmic choices, as explained in the supplement. Here, we focus on conveying the general MSM procedure. 
    
We build the sets $S_{t,i}$ using $k$-means clustering of the data using the \texttt{scikit-learn} package \cite{Pedregosa2011scikit}. As input to $k$-means, we use a vector of feature $\Phi$ consisting of time-delays of $U$:
%\begin{linenomath*}
\begin{align}
	\Phi(\mathbf{X}(t))=[U(\mathbf{X}(t)), U(\mathbf{X}(t-1)), \hdots, U(\mathbf{X}(t-\delta+1))]
\end{align}
%\end{linenomath*}
where $\delta$ days is the number of retained time-delays. This time-delay embedding encodes additional information about the atmospheric state, enabling a model based just on the zonal mean wind at 10 hPa. Heuristically, the embedding captures approximate time-derivatives up to order $\delta$-1. The technique has precedent in climate science \cite{Ghil2002advanced}, and a growing body of theoretical and empirical evidence supports the use of time-delay coordinates as reliable features for encoding dynamical attractors \cite{Takens1981detecting,Kamb2020timedelay,Broomhead1986extracting,Giannakis2012nlsa,Brunton2017chaos,Thiede2019galerkin,Strahan2021long}. We have also experimented with richer feature spaces including EOFs of geopotential height, but found it unnecessary.  

We find that any $\delta$ from 2 to 10 and any number of clusters (denoted $M_t$) from 50 to 150 gives similar results. In Fig.~\ref{fig:rates} we display results of a single representative choice of $\delta=5$ days and $M_t=150$, along with a shaded 95\% confidence interval derived from the pivotal bootstrap procedure \cite{Wasserman2004all} with 20 independent resamplings of the data (but without replacement). The supplement further explains how we selected these parameters to simultaneously optimize the MSM's fidelity and robustness on a simple performance benchmark. We emphasize that these clusters are not supposed to identify metastable weather regimes in the tradition of, e.g., \cite{Michelangeli1995weather}; rather, they are a discretization of state space meant to represent continuous functions over that space, encoding gradual progress towards an SSW event. 

Given the clusters $\{S_{t,j}\}$ and the transition matrices $\{P_{t,t+1}\}$, we can calculate the rate and seasonal distribution of SSW events with the following procedure. 

\begin{enumerate}
        \item Let $B$ denote the set of ``weak-vortex'' clusters: all $(t,j)$ such that $T_0\leq t\leq T_1$ and the majority of data points in $S_{t,j}$ have $U<U^{\text{(th)}}$. Let $A$ denote the set of ``non-winter'' clusters: all $(t,j)$ such that $t<T_0$ or $t>T_1$. With this setup, an SSW event is a \emph{transition from $A$ to $B$}. % TODO explain why "SSW" and "non-SSW" are not appropriate terms here
	\item Compute the \emph{committor probability},
        %\begin{linenomath*}
	\begin{align}
		q^+_t(j)&=\mathbb{P}\Big\{U(s)\leq U^\text{(th)}\text{ for some }s\in[t,T_1]\ \Big|\ \mathbf{X}(t)\in S_{t,j}\Big\},
	\end{align}
        %\end{linenomath*}
	by solving the following terminal/boundary-value problem. By definition, $q^+_{T_1+1}(j)=0$ for all clusters $j$ at the end of winter, while $q^+_t(j)=1$ for all $(t,j)\in B$. Stepping backward through time, we have a recursion relation:
        %\begin{linenomath*}
	\begin{align}
		\label{eq:q_recursion}
		q^+_t(j)&=\sum_{k=1}^{M_{t+1}}P_{t,t+1}(j,k)q^+_{t+1}(k).
	\end{align}
        %\end{linenomath*}
	In words, for a vortex that is initially strong today $(t)$ to break down by Feb. 28 ($T_1$), it must break down sometime between tomorrow ($t+1$) and $T_1$. Hence $q^+_t(j)$ is a weighted combination of $q^+_{t+1}(k)$ for all possible scenarios $k$ for tomorrow. This equation is simply the Kolmogorov Backward Equation in discrete form \cite{E2019applied}. In this light, viewing Eq.~\eqref{eq:q_recursion} as a discretized partial differential equation, the clusters $\{S_{t,j}\}$ can be seen as members of a finite element basis and $P_{t,t+1}(i,j)$ as stiffness matrices. Indeed, here we use an MSM as a ``dynamical Galerkin approximation'', a basis expansion approach to computing forecast quantities like the committor probability from short trajectory data that was originally developed for chemistry applications \cite{Thiede2019galerkin,Strahan2021long} and has recently been applied to climate dynamics \cite{Finkel2021learning,Finkel2022data,Jacques-Duma:2022:comparison}. 		
	\item Estimate an empirical probability distribution over clusters at the beginning of winter,
        %\begin{linenomath*}
	\begin{align}
		\pi_{T_0}(j)&=\mathbb{P}\{\mathbf{X}(T_0)\in S_{T_0,j}\}
	\end{align}
        %\end{linenomath*}
	However, in practice, the result is not sensitive to the choice of initial probability distribution. This is because $T_0$ is early enough in the winter season that the distribution of $U$ is still narrow (see Fig.~\ref{fig:ssw_data}) and the memory of initial conditions is practically erased by the time of the first SSW. We can also propagate $\pi$ to each day of the season, using the Kolmogorov Forward equation (a.k.a. the Fokker Planck equation) in discrete form:
        %\begin{linenomath*}
	\begin{align}
		\pi_{t+1}(k)&=\sum_{j=1}^{M_t}\pi_t(j)P_{t,t+1}(j,k)
	\end{align}
        %\end{linenomath*}
    \item Compute the rate as the average of committor probabilities on the first day of the SSW season, weighted by the probability distribution $\pi_{T_0}$:
        %\begin{linenomath*}
	\begin{align}
		\label{eq:rate_formula}
		R&=\sum_{j=1}^{M_{T_0}}q^+_{T_0}(j)\pi_{T_0}(j)
	\end{align}
        %\end{linenomath*}
	In words, the probability of SSW in a random year is the sum of probabilities from every possible initial condition, weighted by the probability of that initial condition. Fig.~\ref{fig:rates} shows in purple the rate according to the MSM , which matches remarkably well with the flux-counting method. Error bars indicate the 95\% confidence interval, obtained with the same bootstrapping procedure that we used for flux-counting. In particular, the \emph{entire clustering} procedure is repeated for each 10-year subset of data. 
	
    \item Compute the seasonal distribution by decomposing the rate over all possible entrance times to $B$, rather than exit points (i.e., initial conditions) from $A$: 
        %\begin{linenomath*}
	\begin{align}
		R&=\sum_{t=T_0-1}^{T_1}\sum_{j=1}^{M_t}\sum_{k=1}^{M_{t+1}}q_t^-(j)P_{t,t+1}(j,k)\mathbb{I}\{(t+1,k)\in B\}
	\end{align}
        %\end{linenomath*}
	where $q_t^-(j)=\mathbb{P}\{\text{no SSW has occurred yet between $T_0$ and }t|\mathbf{X}(t)=j\}$ is known as the \emph{backward committor}. The backward committor obeys a recursion analogous to that of $q_t^+$, but moving backward through time and with a time-reversed transition matrix:
        %\begin{linenomath*}
	\begin{align}
		q_{t+1}^-(k)&=
		\begin{cases}
			\sum_{j=1}^{M_t}P_{t,t+1}(j,k)\frac{\pi_t(j)}{\pi_{t+1}(k)}q_t^-(j) & (t,j)\notin A\cup B \\
			0 & (t,j) \in B \\
			1 & (t,j) \in A
		\end{cases}
	\end{align}
        %\end{linenomath*}
	The purple histogram in Fig.~\ref{fig:rates}b-e is given by the individual summands (in groups of 7, according to the bin width of 7 days). 

\end{enumerate}

The committor, defined in step 2 above, measures probabilistic progress towards an SSW event (how likely). To measure \emph{temporal} progress (how soon), we further define the \emph{hitting time} as
%\begin{linenomath*}
    \begin{align}
        \tau^+_t=\min\{s\geq 0:(t+s,\mathbf{X}(t+s))\in B\}
    \end{align}
%\end{linenomath*}
This is a random variable that tells you the timing of the SSW, depending on the realization of $\mathbf{X}$. We compute two summary statistics of this random variable. First, its cumulative probability mass function $\mathbb{P}\{\tau_t^+<\sigma\}$ is a \emph{time-limited} version of the committor, which we use to validate our choice of MSM parameters (see the supplement). Second, its average value, conditional on the vortex actually breaking down the same winter, is called the \emph{expected lead time}:
%\begin{linenomath*}
\begin{align}
    \eta_t^+=\mathbb{E}\big[\tau_t^+|t+\tau_t^+\leq T_1\big]
\end{align}
%\end{linenomath*}
This is another useful summary statistic to quantify how far away the system is from an SSW event. We displayed a similar quantity in \cite{Finkel2021learning,Finkel2022data} in the context of an idealized model. The expected lead time can also be computed by recursion with the MSM, but the formula is slightly more involved and left to the supplement.

Let us take a brief aside to reference some mathematical context for the method above. 
The general framework that we have used to combine committor probabilities to compute rates and other steady-state statistics of rare transitions is \emph{transition path theory} (TPT)
% , which describes the steady-state statistics of rare transition events
~\cite{VandenEijnden2014transition}. TPT has been applied to molecular dynamics \cite{Noe2009constructing,Meng2016cSrcTPT,Strahan2021long,Antoszewski2021insulinhex}, atmospheric and oceanic sciences \cite{Finkel2020paths,Finkel2022data,Miron2021transition,Miron2022transition} and social sciences \cite{Helfmann2021statistical}. Though TPT is typically formulated in a time-homogeneous setting, here we have built in explicit time-dependence to deal with the seasonal cycle, similarly to \cite{Helfmann2020extending}. 

Our MSM-based approximation of the committor probability is similar in spirit to analogue forecasting \cite{Dool1989new}, which is enjoying a renaissance with novel data-driven techniques, especially for characterizing extreme weather \cite{Chattopadhyay2020analog,Lucente2022coupling}. Dynamical Galerkin approximation (using a basis different than the one used here) and a short trajectory variant of analogue forecasting are tested on several benchmark problems in  \cite{Jacques-Duma:2022:comparison}.
Formally, the transition operator encoded by the matrix in \eqref{eq:transition_matrix} is related to linear inverse models 
% \cite<LIMs;>[]{Penland1995optimal}
\cite{Penland1995optimal}, which have also been used to predict subseasonal extremes \cite{Tseng2021mapping}. Both MSMs and linear inverse models involve finite-dimensional approximations of the transition operator (or Koopman operator for deterministic dynamics) \cite{Mezic2005analysis,Mezic2005spectral,Klus2018datadriven}. 

\section{Results}

\subsection{Rate estimates}
\label{sec:comparing_rates}

Fig.~\ref{fig:rates} compares rate estimates from the MSM and flux-counting methods against the reanalysis rates. We include the ERA-5 estimator based on just 1996-2015 to overlap with the S2S period. For the mild thresholds of $U^{\mathrm{(th)}}=0,-4$ m/s, corresponding to return times of 2-3 years, the MSM and flux-counting estimates agree with both short- and long-term reanalysis estimates. Moving to moderate thresholds of $U^{\mathrm{(th)}}\sim-28$ m/s, the MSM and flux-counting rates track somewhat closer with the 61-year estimate (orange), which has slightly lower rates across the board. The S2S data were initialized from the 20-year time period corresponding to the blue curve, but the S2S hindcasts recover the longer-term climatology, despite the (slightly) greater SSW frequency from the period in which they were initialized. 

One can think of these SSW frequencies as the \emph{climatology according to the Integrated Forecast System}, given the boundary conditions of the 1996-2015 period.  At least in the ``model world'' of the IFS, it does not appear that a differences in atmospheric boundary conditions (e.g., sea surface temperatures) caused a systematic increase in SSW frequency between 1996 and 2015; rather the observed increase in SSWs was luck of the draw. It is possible, however, that systematic model error could  be obscuring the systematic differences suggested by \cite{Reichler2012stratospheric} and \cite{DimdoreMiles2021origins}. 

At all levels of $U^{\mathrm{(th)}}$, but especially in the negative extremes, the confidence intervals from the two S2S estimates are smaller than those from the ERA5 estimates, thanks to the large amount of S2S data that the MSM and flux-counting methods can exploit.  The agreement between MSM and ERA-5 on common events gives us more confidence to trust the MSM on less common events in the negative-$U^{(\mathrm{th})}$ tail, where ERA-5 data are too sparse to give a meaningful rate estimate.  Both the direct counting and MSM approaches suggest the potential for events where the vortex becomes so disrupted it spins -40 m/s (stronger than average, but in the opposite direction), albeit only once or twice in a millennium. 

Several recent studies have performed the same task of filling out a sparse climate distribution using models \cite{Horan2017modeling,Kelder2020unseen}, but with uninterrupted long runs of a global climate model. The techniques we have introduced---MSM and flux-counting---offer a novel way to estimate such quantities from short trajectories only, without access to a centennial-scale run of the IFS which the standard estimation method would require. We believe the higher resolution IFS is also more appropriate for capturing the most extreme SSWs.

\subsection{Seasonal distribution}

Fig.~\ref{fig:rates}b-e illustrate that S2S data also offers an advantage for describing the seasonal distribution of SSW events, an inherently noisier statistic than the full-winter rate estimate because one has to split the data into finer categories. The S2S-derived histograms, in black and purple, are able to bring out seasonal structure that is ambiguous in the reanalysis data directly. For $U^{\mathrm{(th)}}=0$ m/s, there is a gradual increase in SSW frequency from November to January, followed by a plateau in February, consistent with prior studies of seasonality at monthly resolution \cite{Charlton2007part1} and supporting the late winter maximum found by \cite{Horan2017modeling}. 

As $U^{\mathrm{(th)}}$ becomes more negative, the reanalysis histograms dwindle and degenerate into a few isolated spikes, whereas the S2S histograms become intriguingly bimodal, but retain their smoothness. The S2S histograms show a persisting SSW occurrence throughout February after the second peak, a feature that is also faintly present in the longer reanalysis period (1959-2019), but not at all in the shorter reanalysis period (1996-2015) from which S2S was initialized. Again, the IFS recovers features of the longer-term climatology.  

The January/February peak is documented in the literature, e.g., by \cite{Horan2017modeling}, who diagnosed the peak as a balance between two time-varying signals: the background strength of the polar vortex, and the vertical flux of wave activity capable of disturbing the vortex. The bimodal structure seen in S2S has also been found tentatively in prior studies with both reanalysis and models \cite{Horan2017modeling,Ayarzaguena2019representation}, and more robustly in other features of the boreal winter, e.g., the midwinter suppression of Pacific storm activity \cite{Nakamura1992midwinter}. We speculate that the early peak represents Canadian warmings \cite{Meriwether2004mesosphere}, which our result suggests may deserve a more decisive classification. Seasonal differences are associated with dynamical differences in SSW events. For example, ``Canadian warmings'' shift the Aleutian high and occur earlier in the winter \cite{Butler2015defining}. Categorizing SSWs by their seasonality may reveal preferred timings that indicate when and why the polar vortex is most vulnerable \cite{Horan2017modeling}.

\subsection{Statistical predictors of SSWs}

\begin{figure}%[tbhp]
	\centering
	\includegraphics[trim={0cm 0cm 0cm 0cm},clip,width=.99\linewidth]{"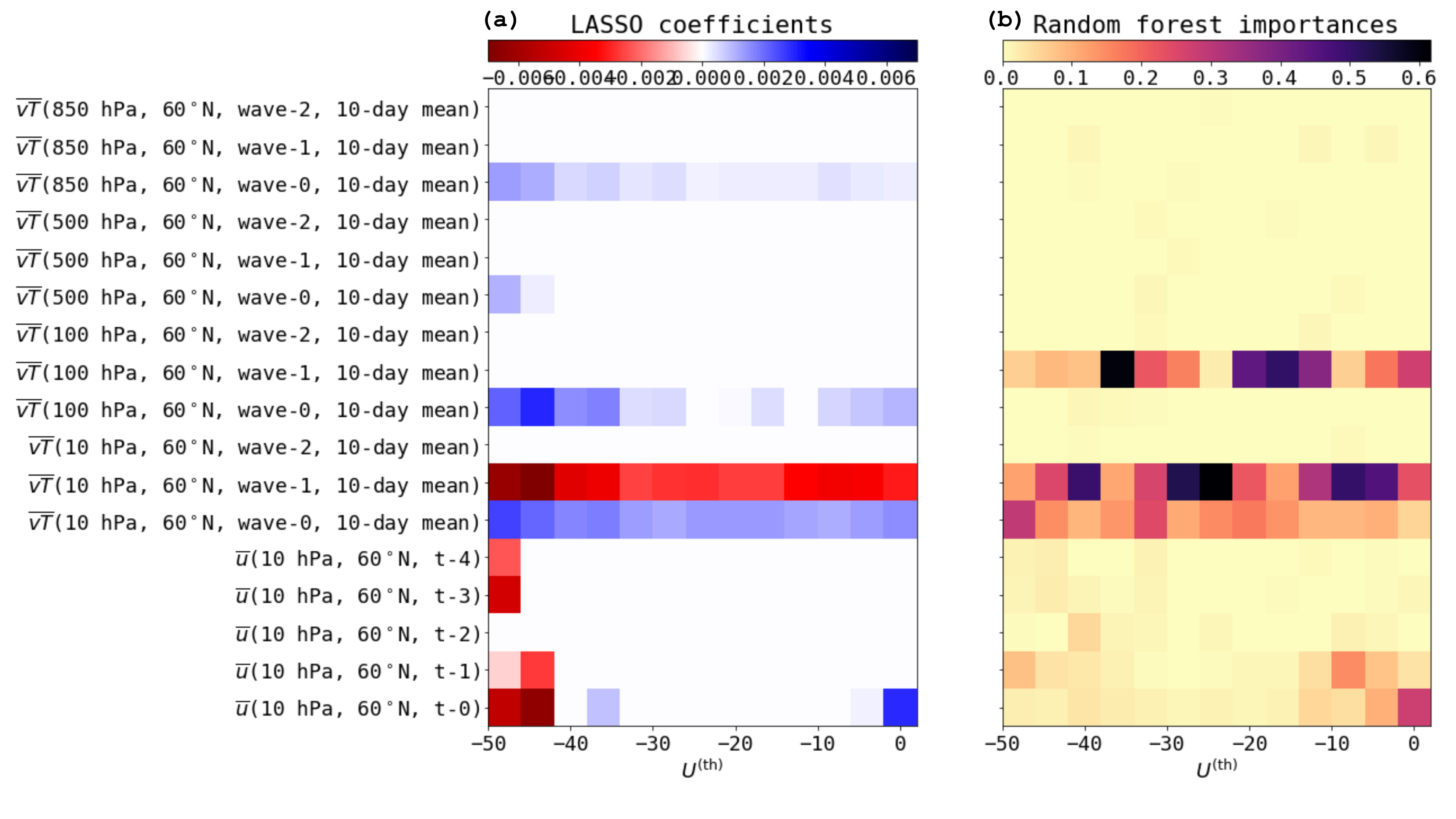"}
	\caption{\textbf{Sparse regression results}. Heat maps show the importance of each feature (listed on the vertical axis) for predicting the expected lead time $\eta^+$ at a range of zonal wind thresholds $U^{(\mathrm{th})}$ (listed along the horizontal axis). The left-hand heat map shows the Gini importances from random forest regression, and the right panel shows LASSO coefficients. We also used $\overline{u}$ at lower levels than 10 hPa as input features, but found none of them to have any importance, and so omitted them from the figure. }
	\label{fig:regression_alluth}
\end{figure}

Estimates of long return times alone do not provide physical insight into the mechanisms driving the event. The committor probability and expected lead time estimates provided by the MSM approach encode information on the dynamics and predictability of SSW events, and on extreme events in general. These quantities cannot be computed by the flux-counting approach. A number of recent articles have pursued committor probabilities as windows into transitional dynamics, e.g., \cite{Miloshevich2022probabilistic} for European heat waves and \cite{Frishman2022dynamical} for the spread of turbulence in a pipe. On SSWs specifically, our own previous studies with a simple SSW model \cite{Finkel2021learning,Finkel2022data} found through sparse regression that a small set of physical variables could explain key variability in the committor. 

Here we analyze the S2S dataset in a similar way, using sparse regression to reveal the main determinants of the committor $q_t^+$ (how likely is an SSW to occur?) and expected lead time $\eta_t^+$ (if it does occur, how soon?), among a large collection of candidate variables including zonal-mean zonal winds and meridional eddy heat fluxes at various time delays, altitudes, and wavenumbers (listed on the left of Fig.~\ref{fig:regression_alluth}). We have performed two kinds of regression: linear regression with a sparsity-promoting $L1$ penalty of $0.1$, also known as LASSO \cite{Tibshirani1996regression}, and random forest regression \cite{Hastie2009elements} with 10 trees of depth 3. Both algorithms, as implemented using \texttt{scikit-learn}, provide not only predictions of the output variable but also notions of relevance for each input feature: nonzero coefficients in the case of LASSO, and Gini importances in the case of the random forest \cite{Pedregosa2011scikit}. These relevances are of greater interest to us than the raw skill of the regression.  

We focus on the early part of the SSW season to connect the results with the rate formula~\eqref{eq:rate_formula}. The target variable for regression is $\log(\eta_t^+)$, which guarantees the predicted $\eta_t^+$ is positive and also emphasizes variability in small values of $\eta_t^+$ (when an SSW is close) rather than large values (when an SSW is distant). The training data consist of trajectory snapshots as inputs and MSM-labeled $\eta_t^+$ values as outputs. We include only those snapshots between Nov. 1 and Nov. 30, and strictly outside of sets $A$ and $B$, where $0<q_t^+<1$. (We also regressed on $q_t^+$ and discovered similar but more subtle patterns of importance; for brevity we show results only for $\eta_t^+$.)

Fig.~\ref{fig:regression_alluth} summarizes the results of regression across all $U^{\mathrm{(th)}}$ thresholds. Random forest importances are always nonnegative and represented on a yellow-black color scale, while regression coefficients are signed and represented on a red-blue color scale. Red is associated with a weaker vortex, meaning a shorter lead time. The correlation coefficient $R^2$ remains between 0.4 and 0.6 for both methods across all thresholds, indicating that these regressions are imperfect expressions of the expected lead time, but do explain a significant part of the variance. 

The models illuminate some interesting patterns, some obvious and some surprising. \emph{A priori}, one expects $\overline{u}$(10 hPa, 60$^\circ$N, $t$) itself [the bottom listed feature, abbreviated $U(t)$] to dominate the regression, since it defines the event. This is true at a mild threshold of $U^{(\mathrm{th})}$---stronger zonal wind means longer expected lead time, according to the positive coefficient in the panel's lower right corner---but for more extreme thresholds, it is actually the \emph{time-delayed} zonal wind $U(t-1),\hdots,U(t-4)$ that is more relevant. Furthermore, the corresponding LASSO coefficients are negative, suggesting that the decrease over time of $U$ is more important than its value today. At the most extreme thresholds, it even appears that \emph{strong} $U(t)$ portends a sooner vortex collapse, suggesting that the most extreme SSW events (those reaching the most \emph{negative} zonal wind) follow from precursor states with anomalously \emph{strong} zonal wind. %This would be consistent with Fig.~\ref{fig:ssw_data}, which shows the 2009 SSW descend from a near-record high $U$ to a record low $U$. However, this is only a cautious speculation, as the extreme thresholds are still very data-limited.

Another important set of features is the 10-day averaged meridional heat flux $\overline{vT}$ averaged over 45-75$^\circ$N, although LASSO and Random Forest regressions emphasize different altitudes and wavenumber components. Both methods agree that the 10 hPa heat flux at wavenumbers 0 and 1 exert strong and competing (statistical) influences on expected lead time: a stronger wavenumber-0 component ($\overline{v}\overline{T}$) means vortex collapse is farther away, while a stronger wavenumber-1 component means vortex collapse is sooner. At lower levels of the atmosphere, the eddy heat fluxes exert significant but diminishing influences, although they remain important for the most extreme SSW events (at least according to LASSO).

\begin{figure}%[tbhp]
	\centering
	\includegraphics[trim={0cm 0cm 14cm 0cm},clip,width=.75\linewidth]{"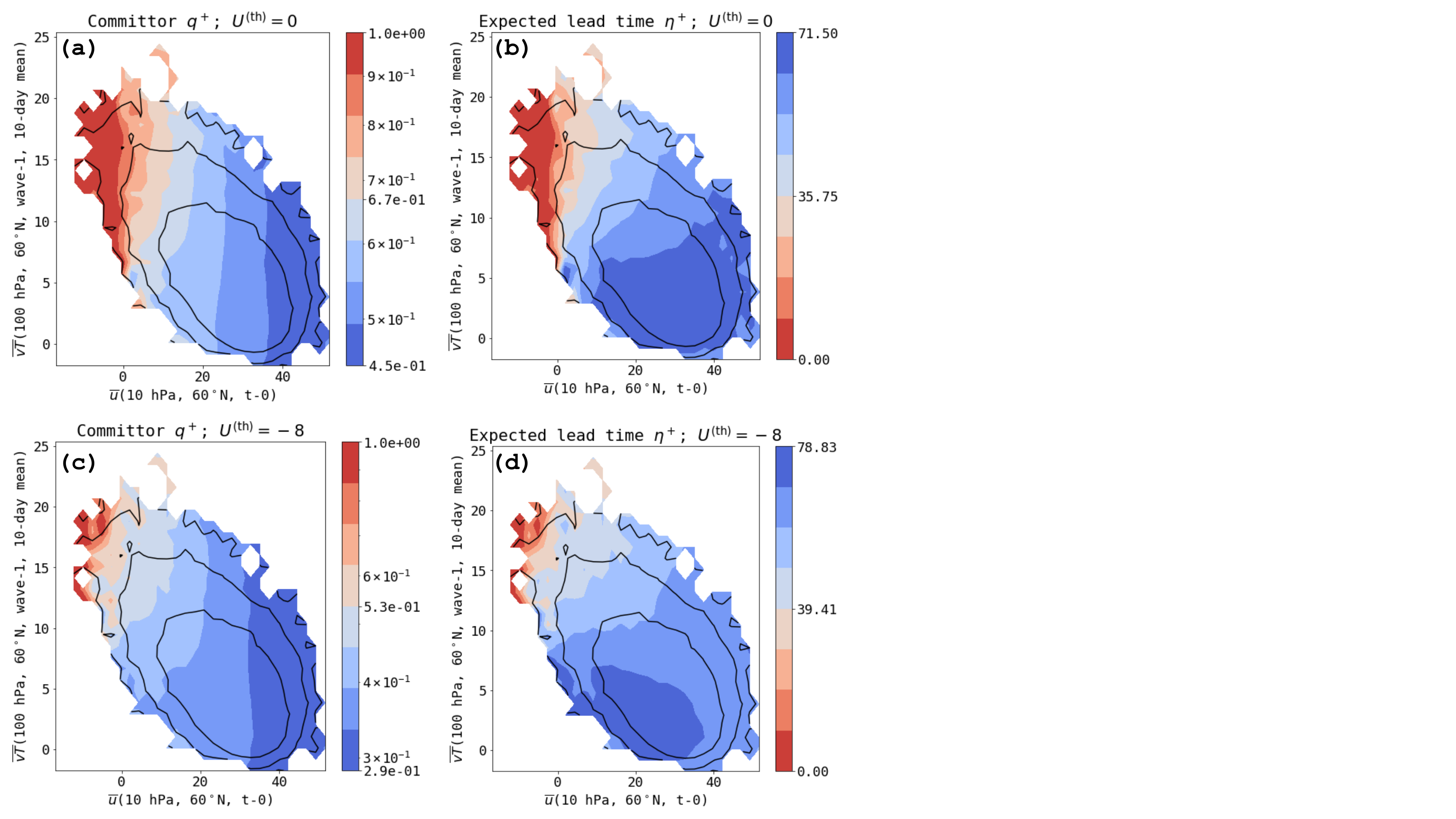"}
	\caption{\textbf{Committor and expected lead time}. Aggregating all S2S data from Nov. 1 to Nov. 30, this figure displays the committor (left column) and expected lead time (right column) in shading, as well as the climatological probability density $\pi$ in black contours, on a logarithmic scale ($\pi$ has arbitrary units, normalized to have unit integral over state space). Two zonal wind thresholds are considered: 0 m/s (top), and $-8$ m/s (middle) (bottom). All results are derived from the MSM. We show $\eta^+$ and $q^+$ as functions of two variables only: $\overline{u}$(10 hPa, 60$^\circ$N, $t$) and meridional heat flux averaged over 10 days between 45$^\circ$N and 75$^\circ$N at 100 hPa and wavenumber 1. The remaining variables are averaged out and weighted by $\pi$ in this display.}
	\label{fig:projections_2d}
\end{figure}

How can we make sense of all these correlations? One simple method of visualization is to plot the committor and lead time as approximate functions of two variables (averaging over remaining variables). The regression results present us with many possible pairs of important variables. Here we select just one pair: 
zonal-mean zonal wind at 10 hPa, and wavenumber 1 meridional heat flux at 100 hPa, averaged over the preceding 10 days. The latter feature is assigned high importance by the random forest, though not by LASSO, and is especially interesting as a signal coming from a lower altitude than 10 hPa, possibly related to the two-way influence characteristic of coupled troposphere/stratosphere dynamics. Fig.~\ref{fig:projections_2d} displays the committor (left column) and expected lead time (right column) as a function of these two variables at two thresholds: $U^{(\mathrm{th})}=0$ m/s (top), and $-8$ m/s (bottom). Contours of the climatological probability density $\pi$ signal which regions are more frequently visited and which ones are rare. We only average over the first month of the SSW season, Nov. 1-Nov. 30, to represent a map of possible ``initial conditions'' for the winter vortex evolution. 

The orientation of contours in phase space reveals a pattern of influence that would be hard to intuit from the regression coefficients alone. At $U^{(\mathrm{th})}=0$, the $q^+$ contours run almost perpendicular to the $U$ axis, confirming that the zonal wind itself primarily determines how likely an SSW is for the coming season. But the $\eta^+$ contours tell a different side of the story: at stronger $U$, the contours progressively tilt away from vertical towards horizontal, indicating that the time until SSW depends strongly on heat flux---at least in the regime of strong $U$, from where an SSW is unlikely to begin with. The influence of heat flux grows more significant as the threshold $U^{(\mathrm{th})}$ is lowered in row 2 of the figure, even for the committor. We infer a general pattern: the 10-hPa zonal wind strength in November determines how likely an SSW is for the coming winter, but when it is rather unlikely, the lower-stratospheric wavenumber-1 heat flux determines when the SSW will happen. 

What the phase space images reveal most of all is that $q^+$ and $\eta^+$ are nonlinear functions: the influence of a variable depends on the state of all the other variables. Nonlinear regression methods, such as random forests, are therefore crucial to uncover a complete description.

% Neural networks are another possibility, as demonstrated by \cite{Miloshevich2022probabilistic} who use convolutional neural networks to learn committor functions for heat waves, and thus diagnose short- and long-term meteorological predictors. Data-hungry neural networks stand to benefit from datasets such as S2S, whose short-duration limitations can be overcome with, e.g., the MSM method we have shown. 

\section{Discussion}
\label{sec:discussion}

By comparing S2S results with reanalysis, we are measuring the composition of potentially three separate error sources: (i) forecast model error, (ii) non-stationarity of the climate \emph{with respect to SSW events} over the reanalysis period, and (iii) numerical errors, both statistical (from the finite sample size) and systematic (from the projection of forecast functions onto a finite basis in the case of the MSM). We briefly address each error source in turn. 

The S2S trajectories were realized only in simulation, not in the physical world. Accordingly, our S2S estimates apply strictly to the climatology of the 2017 IFS, a statistical ensemble that could (at least in principle) be concretely realized by running the model uninterrupted for millennia, with external climatic parameters sampled from their variability in the 20-year time window of 1996-2015. Long, equilibrated simulations have been performed with coarser models by, e.g., \cite{Kelder2020unseen} to assess UK flood risk (the so-called ``UNSEEN'' method), and by \cite{Horan2017modeling} to assess SSW frequencies, but this is not practical given the constraints and mission of the ECMWF IFS. The S2S dataset is an ensemble of opportunity. It was created to compare the skill of different forecast systems on S2S timescales, not at all for the purpose of establishing a climatology of SSWs. 

And while the IFS model has proven outstanding in its medium-range forecast skill \cite{Vitart2013evolution,Kim2014predictability,Vitart2018s2s}, it was designed for short forecasts. It is not clear how it would behave if allowed to run for hundreds of years as a climate model, which requires careful attention to the boundary condition and conservation issues. Even if the climate were to remain stationary with its 1996-2015 parameters, numerical and model errors would inject some bias into the equilibrated simulation. Repeatedly initializing S2S forecasts with reanalysis ensures a realistic background climatology, and allows us to rely on the IFS strictly for the short-term integrations that it was designed for. Our method may be used as a diagnostic tool to compare different models against each other, with specific attention paid to their rare event rates. A useful extension of this work would be to repeat the analysis on multiple data streams from all 11 forecasting centers worldwide that contribute to the S2S project, as a different way to compare different models' ability to represent extremes. 

Boreal SSWs provide an ideal demonstration of our method, providing both moderately and extremely rare events. A natural and intriguing future application is the rate of Southern-hemisphere SSW events, in the spirit of \cite{Jucker2021how}. which is postponed to future work for the sake of brevity. The method may be extended to other kinds of extremes as well, though care must be exercised when defining the event (e.g., sets $A$ and $B$) and choosing features in which to do clustering (for the MSM approach), especially in the case of more spatially localized events.

The rate we estimate from the S2S data set is based on 1996-2015 boundary conditions (sea surface temperatures, CO2), and our MSM method assumes the climate was stationary over this period. Our results indicate that according to the 2017 IFS, 1996-2015 conditions were more similar to 1959-2019 than direct counting of SSW events might suggest.  This could mean that the IFS was missing some key climatological variable during that period \cite{DimdoreMiles2021origins}. There is, however, substantial uncertaintuy on the impact of global warming on SSWs, even under 4xCO2 forcing %insufficient evidence on the anthropogenic influence on SSW to reject the hypothesis of stationarity 
\cite{Ayarzaguena2020uncertainty}. %The potential impact of global warming on SSWs is an important open question. %, although we have restricted this paper to stationary statistics in order to focus on the method. However, 
By repeating our analysis on different historical periods, or simulations initialized from climate model integrations under different forcing, one could discern a more decisive signal of forced changes than would be available from raw data. Moreover, the expression for the SSW rate~\eqref{eq:rate_formula} as a ``dot product'' between a committor and a climatological probability density would allow us to decompose small changes in SSW frequency as changes in these two components separately. A changing probability density $\pi$ would reflect changes in the slow background conditions, whereas a changing committor $q^+$ would reflect a change in the system dynamics. 

Error source (iii) is the most open to scrutiny and improvement. We have used the short S2S hindcasts directly to validate our parameter choices for the MSM (see the supplement). In a sequence of preceding papers \cite{Finkel2021learning,Finkel2022data}, we have benchmarked the performance of the method on a highly idealized SSW model due to \cite{Holton1976stratospheric}. Nevertheless, large-scale atmospheric models are a mostly-unexplored frontier for this class of methods.
% , and there is much improvement to be had, for example, with more advanced dimensionality reduction methods.  

Our method exceeds what is possible directly from reanalysis, but we are not yet fully liberated from observations: every S2S trajectory is initialized near reanalysis, and it only has 46 days to explore state space before terminating. This fundamentally limits how far we can explore the tail of the SSW distribution. In other words, the real climate system sets the \emph{sampling distribution} which is a flexible but important component in rare event estimation problems \cite{Thiede2019galerkin,Strahan2021long,Finkel2021learning}. With an executable model, we could initialize secondary and tertiary generations of short trajectories to push into more negative-$U$ territory and maintain statistical power for increasingly extreme SSW events. This is the essence of many rare-event sampling algorithms, such as those reviewed in \cite{Bouchet2019rare} and \cite{Sapsis2021statistics}. For example, a splitting large-deviation algorithm was used in \cite{Ragone2018heatwaves} to sample extreme European heat waves and estimate their return times. Quantile diffusion Monte Carlo was used in \cite{Webber2019practical} to simulate intense hurricanes, and in \cite{Abbot2021rare} to estimate the probability of extreme orbital variations of Mercury. A natural extension of these various techniques would combine elements of active rare event sampling with committor estimation via MSMs. Early developments of such a coupling procedure are presented in \cite{Lucente2022coupling}.

\section{Conclusion}
\label{sec:conclusion}
Extreme weather events present a fundamental challenge to Earth system modeling. Very long simulations are needed to generate sufficiently many extreme events to reduce statistical error, but high-fidelity models are needed to simulate those events accurately. Conventionally, no single model can provide both, due to computational costs. Here, we have demonstrated an alternative approach that leverages \emph{ensembles} of \emph{short}, high-fidelity weather model forecasts to calculate extreme weather statistics, with specific application to sudden stratospheric warming (SSW). By exploiting the huge database of forecasts stored in the subseasonal-to-seasonal (S2S) database \cite{Vitart2017subseasonal}, we have obtained estimates of the rate and seasonal distribution of extreme SSW events. From just 20 years of data, we obtain probability estimates of events with a 500 year return time, which are so extreme that the vortex is as strong in the easterly direction as its typical westerly climatology. These events have never been observed historically, but can be pieced together using our analysis method. 

Our method uses data to estimate the dynamics on a subspace relevant for SSW, namely the polar vortex strength as measured by zonal-mean zonal wind. This single observable, augmented by time-delay embedding, gives a simple set of coordinates sufficient to estimate rate and seasonal distributions. Our demonstration opens the door to address many other data-limited questions of basic physical interest. For instance, a high-resolution model could be used in ensemble forecast mode, but initialized around a decade at the end of this century provided by a climate model, to understand the impact of global warming on extremes.

\section{Open Research}
Our analysis is based on publicly available datasets from the European Center for Medium-Range Weather Forecasts \cite{ECMWFint} and from associated Copernicus Climate Data Store \cite{Copernicus}. Python scripts to download the necessary data and reproduce the paper's analysis will be made available in a public Zenodo repository at publication time.

%%%%%%%%%%%%%%%%%%%%%%%%%%%%%%%%%%%%%%%%%%%%%%%

\section*{Acknowledgments}
We extend special thanks to Andrew Charlton-Perez, who suggested the S2S dataset as a case study for the methodology, and Simon Lee, who helped familiarize us with the data. We thank Amy Butler for help with the reanalysis. Our collaborators at the University of Chicago, including Aaron Dinner, John Strahan, and Chatipat Lorpaiboon, offered helpful methodological advice. Computations for this project were performed on the Greene cluster at New York University. 

J.F. was supported by the U.S. DOE, Office of Science, Office of Advanced Scientific Computing Research, Department of Energy Computational Science Graduate Fellowship under Award Number DE-SC0019323 at the time of writing. J. F. also acknowledges continuing support from the MIT Climate Grand Challenge on Weather and Climate Extremes. E.P.G. acknowledges support from the NSF through award OAC-2004572. J.W. acknowledges support from the NSF through award DMS-2054306 and from the Advanced Scientific Computing Research Program within
the DOE Office of Science through award DE-SC0020427.

\clearpage

\appendix

\section{Supporting information}

Our work relies completely on publicly available datasets of reanalysis and hindcasts, which we describe in the subsequent section. We then lay out the numerical procedure to compute rates and seasonal distributions using transition path theory (TPT). We then present the formulas used to display results in the main text. Finally, we document the method used to select parameters. A software implementation will be made available in a public repository at the time of publication.

%\clearpage

%Delete all unused file types below. Copy/paste for multiples of each file type as needed.
\noindent\textbf{Dataset description}

We use two different datasets for this study.
\begin{itemize}
	\item ERA-5: reanalysis product from ECMWF \cite{Hersbach2019era5}, spanning 1959-2019. We used daily averaged temperature, geopotential height, zonal and meridional wind at 10, 100, 500, and 850 hPa levels at a resolution of 2.5$^\circ$. ERA-5 was downloaded from the Copernicus Data Store \url{https://cds.climate.copernicus.eu/}
	\item S2S: perturbed reforecast (hindcast) ensembles from the 2017 model version of the ECMWF IFS, launched every Monday and Thursday, from fall 1996 through spring 2016, at the same resolution as ERA5 above. Each hindcast ensemble has 10 perturbed members which run for 47 day, including the initialization date. We used the same time and space resolution as with ERA5. S2S was downloaded from the ECMWF data portal \url{https://ecmwf.int}.
\end{itemize}

The S2S dataset can be summarized as follows:
\begin{align}
	\text{S2S dataset}=\{\mathbf{X}_i(t): i=1,\hdots,N\} 
\end{align}
Here, $\mathbf{X}$ denotes the state vector of all relevant meteorological variables. Associated with each calendar day $t$ is a subset of ``active'' indices, $\mathcal{I}(t)$, containing the trajectories launched sometime between $t-46$ and $t$. The same notation can be used for the zonal-mean zonal wind itself, i.e., $U_i(t)$ as a function of $\mathbf{X}_i(t)$.

% ------------ pieces of bullet list -----------

\section{MSM calculations}

The main text explains the computation of the committor, $q_t^+$, and probability distribution, $\pi_t$, for a Markov chain. Here we further explain how to compute statistics of the hitting time, 
\begin{align}
	\tau^+_t=\min\{s\geq 0:(t+s,\mathbf{X}(t+s))\in B\}
\end{align}

\subsubsection{PMF of hitting time}
The probability mass function of $\tau_t^+$ conditioned on an initial set $S_{t,j}$ can be computed with a recursion relation similar to that for $q_t^+$. As a base case, we observe that only way for $\tau_t^+=0$ is for the system to already be in state $B$:
\begin{align}
	\mathbb{P}\{\tau^+_t=0|\mathbf{X}(t)\in j\}=\begin{cases}1&\text{ if }(t,j) \in B \\ 0 & \  				\text{ otherwise}\end{cases}			
\end{align}
For the case $\tau_t^+\geq1$, the hitting time is one plus the hitting time at the next step:
\begin{align}
	\mathbb{P}\{\tau_t^+=s|\mathbf{X}(t)\in j\}=\sum_kP_{t,t+1}(j,k)\mathbb{P}\{\tau_{t+1}^+=s-1|\mathbf{X}(t+1)\in S_{t+1,k}\}
\end{align}
To carry out this recursion, we compute the left-hand side for all $t$ and $j$ with a fixed $s$ before incrementing $s\rightarrow s+1$. After doing this for $0,1,\hdots,s$, we then have a \emph{time-limited} committor,
\begin{align}
	q^+_{t,\sigma}(j)=\mathbb{P}\{\tau_t^+\leq\sigma|\mathbf{X}(t)\in S_{t,j}\}=\sum_{s=0}^\sigma\mathbb{P}\{\tau_t^+=s|\mathbf{X}(t)\in S_{t,j}\},
\end{align}
which is the preferred version of the committor for some other studies, e.g., \cite{Lucente2022coupling}. Below we use the time-limited committor for hyperparameter tuning. 

\subsubsection{Expectation of hitting time}
The most intricate computation is that of the expected lead time $\eta_t^+=\mathbb{E}[\tau_t^+|\tau_t^++t<T_1]$. We do this via the \emph{moment-generating function},
\begin{align}
	M_t(j;\lambda)=\mathbb{E}\big[\exp\big(\lambda\tau_t^+\big)\mathbb{I}\{\tau_t^++t<T_1\}\big|\mathbf{X}(t)\in S_j\big]
\end{align}
Along a single trajectory, we have $\tau_t^+=\tau_{t+1}^++1$ and thus $\exp\big(\lambda\tau_t^+\big)=\exp\big(\lambda\tau_{t+1}^+\big)\exp\big(\lambda\big)$. Using this fact, there is a recursive relationship between $M_t$ and $M_{t+1}$:
\begin{align}
	M_t(j;\lambda)=
	\begin{cases}
	\sum_kP_{t,t+1}(j,k)e^\lambda M_{t+1}(k;\lambda) & (t,j)\notin A\cup B \\
	1 & (t,j) \in B \\
	0 & (t,j) \in A
	\end{cases}
\end{align}
One can observe that $M(t;\lambda=0)$ is the committor itself, $q_t^+$. But to get at $\tau_t^+$, we now must differentiate with respect to $\lambda$:
\begin{align}
	\frac{\partial}{\partial\lambda}M_t(j;\lambda)&=
	\begin{cases}
	\sum_kP_{t,t+1}(j,k)e^\lambda\Big[M_{t+1}(k,\lambda)+\frac{\partial}{\partial\lambda}M_{t+1}(k,\lambda)\Big] & (t,j)\notin A\cup B \\
	0 & (t,j) \in B\\
	0 & (t,j) \in A
	\end{cases}
\end{align}
The lead time can be expressed
\begin{align}
	\eta_t^+(j)=\frac{\mathbb{E}[e^{\lambda\tau_t^+}\mathbb{I}\{\tau_t^++t<T_t\}|\mathbf{X}(t)\in S_j]}{\mathbb{P}\{\tau_t^++t<T_1|\mathbf{X}(t)\in S_j\}}
	=\frac{[\partial M_t(j;\lambda)/\partial\lambda]_{\lambda=0}}{q_t^+(j)}
\end{align}
Therefore, the recursion relation for $\eta_t^+$ is found by setting $\lambda=0$ in the recursion relation for $M_t(j;\lambda)$:
\begin{align}
	\eta_t^+(t,j)&=
	\begin{cases}
	\frac1{q_t^+(j)}\sum_kP_{t,t+1}(j,k)e^\lambda q_{t+1}(k)\big[1+\eta_{t+1}(k)\big] & (t,j)\notin A\cup B \\
	0 & (t,j) \in B \\
	\text{undefined} & (t,j) \in A
	\end{cases}
\end{align}

The formulas for $q^+$ and $\eta^+$ are exact, straightforward to implement, and fast to compute after a Markov chain is constructed. How to choose parameters to construct the chain in the first place is the focus of the next section. 

% ----------- Hyperparameters ---------------
\section{MSM hyperparameter selection}

\begin{figure}%[tbhp]
	\centering
	\includegraphics[trim={0cm 0cm 0cm 0cm},clip,width=.99\linewidth]{"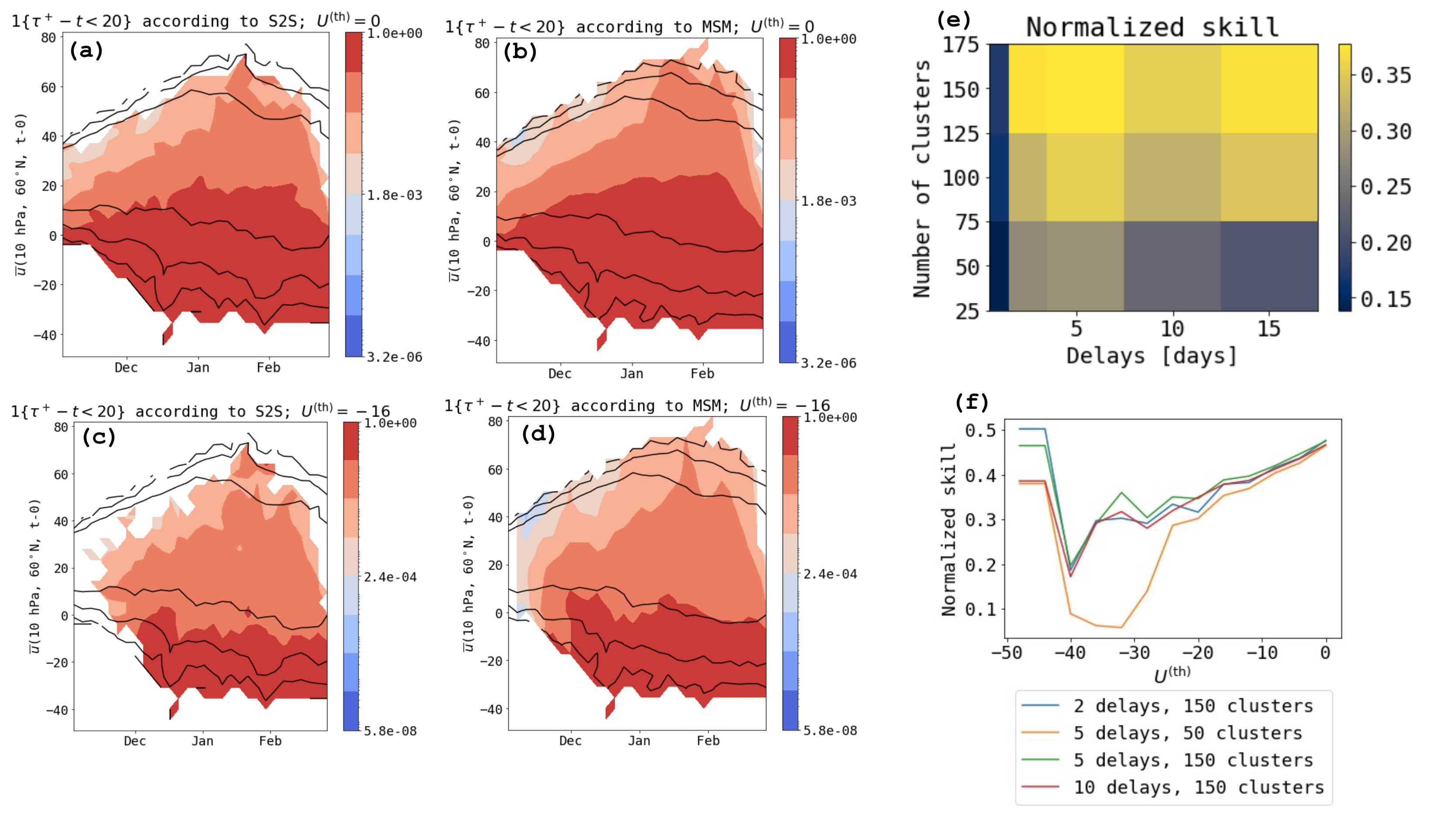"}
	\caption{\textbf{Time-limited committors for hyperparameter selection}. Left two columns: shading indicates time-limited committor probabilities according to counting S2S hindcasts (left) and the MSM (right), at two different thresholds: $U^{\mathrm(th)}=0$ m/s (top) and $-16$ m/s (bottom). Black contours delineate level sets of the climatological probability density, derived from the same two methods. The MSM was constructed using 5 time delays and 150 clusters, a choice informed by systematic evaluation of the MSMs performance. The upper right panel shows the normalized skill of the MSM averaged over zonal wind thresholds, while the lower right panel plots the skill against thresholds for a few selected parameter sets. }
	\label{fig:qplimited}
\end{figure}

The MSM procedure uses two key parameters: the number of time delays $\delta$, and the number of clusters to use in the resulting $(\delta+1)$-dimensional feature space. We optimize this choice by comparing the time-limited committor with $\sigma=20$ according to the MSM ($q^{\mathrm{(MSM)}}$, which takes on continuous values between 0 and 1) with that according to the S2S data directly ($q^{\mathrm{(S2S)}}$, which takes on discrete values of 0 or 1). Forgive the temporary abuse of sub- and superscripts on $q$ in this subsection.  For any ensemble member $\mathbf{X}_i(t)$ at some time $t$ between 0 and 46, the latter can be assessed directly by asking whether $\mathbf{X}_i$ achieves SSW before a time $\sigma$ has elapsed. Fig.~\ref{fig:qplimited} shows $q^+_{t,\sigma}$ as a function of $t$ and $U^{\mathrm{(th)}}$ for $\sigma=20$ days and two different zonal wind thresholds, 0 m/s (top) and $-16$ m/s (bottom). Black contours are level sets of the climatological probability density, $\pi$. The two columns show $q^+_{t,\sigma}$ estimated from S2S directly (left) and the MSM (right), and they match approximately by eye. To quantify their agreement, we use the log-likelihood commonly used for logistic regression, which is equivalent to the (negative) cross entropy of Bernoulli$\big(q^{(\mathrm{MSM})}\big)$ relative to Bernoulli$\big(q^{(\mathrm{S2S})}\big)$:
\begin{align}
	\text{LL}\big(q^{(\mathrm{S2S})},q^{(\mathrm{MSM})}\big)=\text{mean}\Big\{q^{\mathrm{(S2S)}}_i\log\Big(q^{\mathrm{(MSM)}}_i\Big)+\Big(1-q^{\mathrm{(S2S)}}_i\Big)\log\Big(1-q^{\mathrm{(MSM)}}_i\Big)\Big\}.
\end{align}
The committors are evaluated at day 23 of the member $i$, and the mean is taken over the ensemble members $i$ such that $q_i^{\mathrm{(MSM)}}$ is strictly between zero and one on the 23rd day. We choose 23 days because (i) it is beyond the 15-day maximum lag time, before which not all time-lagged features are defined for $\delta=15$, and (ii) it leaves room for 20 days of extra lead time for validation purposes. Other choices in this neighborhood do not affect our results appreciably. 

For extremely rare events---say, with 1\% probability, so that the sample mean $\overline{q}^{(\mathrm{S2S})}=0.01$---there is a class imbalance problem: $q^{(\mathrm{MSM})}$ can achieve a high LL score with a ``climatological'' forecast of $q_i^{(\mathrm{MSM})}\equiv\overline{q}^{(\mathrm{S2S})}$. LL can still be used as a relative score to discriminating between parameter choices at a fixed $\overline{q}^{(\mathrm{S2S})}$, but we wish to choose MSM parameters that are optimal on average across a wide range of $U^{(\mathrm{th})}$ and therefore of $\overline{q}^{(\mathrm{S2S})}$. To compare these scales meaningfully, we convert LL into an absolute scale following \cite{Benedetti2010scoring,Miloshevich2022probabilistic}:
\begin{align}
	\text{Normalized skill}&=
	\frac{\mathrm{LL}\big(q^{(\mathrm{S2S})},q^{(\mathrm{MSM})}\big) - \mathrm{LL}\big(q^{(\mathrm{S2S})},\overline{q}^{(\mathrm{S2S})}\big)} 
	{\mathrm{LL}\big(q^{(\mathrm{S2S})},q^{(\mathrm{S2S})}\big) - \mathrm{LL}\big(q^{(\mathrm{S2S})},\overline{q}^{(\mathrm{S2S})}\big)}
\end{align}
The numerator is the improvement relative to the climatological forecast, and the denominator is the maximum possible improvement, when each $q^{(\text{S2S})}_i$ is predicted exactly. Hence the normalized skill is always less than one, and typically above zero. The right-hand panels of Fig.~\ref{fig:qplimited} display the average normalized skill across thresholds (top) and the normalized skill as a function of threshold (bottom) for a few parameter choices. Increasing the number of clusters seems to help regardless of the number of delays. On the other hand, more delays don't always help; beyond $\delta=5$ days, the dimensionality may be introducing degeneracies. Among the parameter choices considered, the top and bottom panel both suggest that $\delta=5$ delays and $M_t=150$ clusters achieves the best performance among all the MSM choices shown. 

\begin{figure}
	\centering
	\includegraphics[trim={0cm 0cm 0cm 0cm},clip,width=.99\linewidth]{"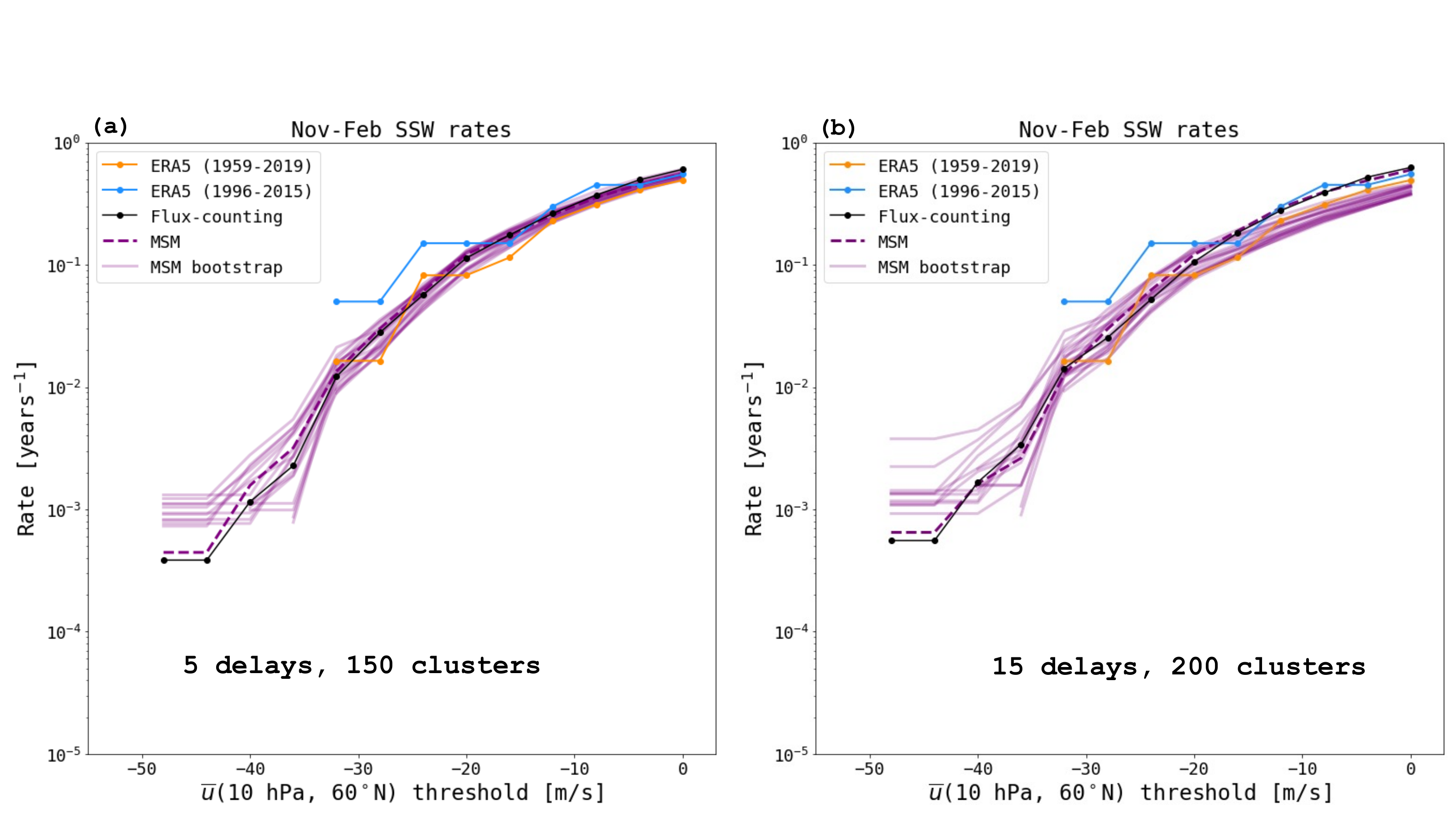"}
	\caption{\textbf{Bootstrap-resampled rates}. Left: reproduction of Fig.~\ref{fig:rates}, plus 20 additional rate curves obtained by resampling the 20 years without into random subsets of size 10, without replacement. Right: same for 15 delays and 200 clusters. The displacement of the bootstrapped curves from the main curve is a symptom of overfitting.}
	\label{fig:bootstrap}
\end{figure}

Fig.~\ref{fig:bootstrap} displays further confirmation that this choice is reasonable and robust. We have repeated the entire MSM pipeline with 20 different random subsets of the years 1996-2015 (sampled without replacement), and plotted the corresponding rates as light purple curves in Fig.~\ref{fig:bootstrap}a. The resampled estimates cluster about the dashed curve, which uses all 20 years simultaneously, for all but the most extreme events. On the other hand, in Fig.~\ref{fig:bootstrap}b, we used 15 delays and 200 clusters, which gave a slightly better normalized skill score (not shown). However, the resampled rates are systematically biased away from the full 20-year estimate, which raises the suspicion of overfitting. For this reason, we stick to the more modest choice of 5 delays and 150 clusters. However, all the qualitative results shown---the committors, expected lead times, and rate curves---remain remarkably consistent with different choices of clusters and delays. This gives us confidence in the MSM results shown in the main text. 

\bibliographystyle{apa-good}
\bibliography{references}

\end{document}